\documentclass[pra,twocolumn,showpacs,superscriptaddress,amssymb,10pt]{revtex4}

\usepackage{graphicx}
\usepackage{dcolumn}
\usepackage{bm}
\usepackage{epsfig}
\usepackage{color}
\usepackage{longtable}
\usepackage{amsmath}
\usepackage{multirow}
\usepackage{tabularx}
\usepackage{siunitx}
\usepackage{soul}
\definecolor{aogreen}{rgb}{0.0, 0.5, 0.0}

\def\ketm#1{  \left\vert  #1   \right\rangle   }

\def\sprm#1#2{  \left\langle #1 \left\vert \right. #2 \right\rangle   }

\def\mem#1#2#3{  \left\langle #1 \left\vert  #2 \right\vert #3 \right\rangle   }

\def\redmem#1#2#3{  \left\langle #1 \left\Vert
                  #2 \right\Vert #3 \right\rangle   }

\definecolor{revisedcolor}{RGB}{0,100,20}

\definecolor{Bcolor}{RGB}{10,200,10}

\definecolor{Jcolor}{RGB}{20,20,200}
\definecolor{Ccolor}{RGB}{200,20,20}
\definecolor{Qcolor}{RGB}{50,200,200}

\DeclareMathOperator*{\SumInt}{%
\mathchoice%
  {\ooalign{$\displaystyle\sum$\cr\hidewidth$\displaystyle\int$\hidewidth\cr}}
  {\ooalign{\raisebox{.14\height}{\scalebox{.7}{$\textstyle\sum$}}\cr\hidewidth$\textstyle\int$\hidewidth\cr}}
  {\ooalign{\raisebox{.2\height}{\scalebox{.6}{$\scriptstyle\sum$}}\cr$\scriptstyle\int$\cr}}
  {\ooalign{\raisebox{.2\height}{\scalebox{.6}{$\scriptstyle\sum$}}\cr$\scriptstyle\int$\cr}}
}

\begin{document}
\include{Bibliography.bib}
\preprint{}
\title{
Polarization effects in the total rate of biharmonic $\omega + 3\omega$ ionization of atoms
}

\author{J.~Hofbrucker}
\affiliation{Helmholtz-Institut Jena, Fr\"o{}belstieg 3, D-07743 Jena, Germany}
\affiliation{GSI Helmholtzzentrum f\"ur Schwerionenforschung GmbH, Planckstrasse 1, D-64291 Darmstadt, Germany}

\author{S.~Ramakrishna}
\affiliation{Helmholtz-Institut Jena, Fr\"o{}belstieg 3, D-07743 Jena, Germany}%
\affiliation{GSI Helmholtzzentrum f\"ur Schwerionenforschung GmbH, Planckstrasse 1, D-64291 Darmstadt, Germany}
\affiliation{Theoretisch-Physikalisches Institut, Friedrich-Schiller-Universit\"at Jena, Max-Wien-Platz 1, D-07743
Jena, Germany}

\author{A.~V.~Volotka}
\affiliation{Department of Physics and Engineering, ITMO University, Kronverkskiy pr. 49, 197101 St. Petersburg, Russia}

\author{S.~Fritzsche}
\affiliation{Helmholtz-Institut Jena, Fr\"o{}belstieg 3, D-07743 Jena, Germany}%
\affiliation{GSI Helmholtzzentrum f\"ur Schwerionenforschung GmbH, Planckstrasse 1, D-64291 Darmstadt, Germany}
\affiliation{Theoretisch-Physikalisches Institut, Friedrich-Schiller-Universit\"at Jena, Max-Wien-Platz 1, D-07743
Jena, Germany}


\begin{abstract}

The total ionization rate of biharmonic ($\omega + 3\omega$) ionization is studied within the independent particle approximation and the third order perturbation theory. Particular attention is paid to how the polarization of the biharmonic light field affects the total rate. The ratios of the biharmonic ionization rates for linearly and circularly polarized beams as well as for corotating and counterrotating elliptically polarized beams are analyzed, and how they depend on the beam parameters, such as photon frequency or phase between $\omega$ and $3\omega$ light beams. We show that the interference of the biharmonic ionization amplitudes determines the dominance of a particular beam polarization over another and that it can be controlled by an appropriate choice of beam parameters. Furthermore, we demonstrate our findings for the ionization of neon $L$ shell electrons.  
\end{abstract}

\newpage
\maketitle


\section{Introduction}
\label{Sec.Intro}

The influence of the polarization of the light beam on the total photoionization cross sections has been a focus of research for decades. While the total cross section for the one-photon ionization of unpolarized atoms is independent of the polarization of the ionizing light, the multi-photon ionization cross section of atoms is typically influenced by its polarization. Based on rather a simple analysis of angular factors, Klarsfeld and Maquet \cite{Klarsfeld1972} predicted that atoms are ionized more efficiently by a circularly polarized light beam than by a linearly polarized beam. Moreover, they also pointed out that the ratio of total multi-photon ionization cross sections for circular versus linear polarization increases rapidly with the order of the process. However, subsequent theoretical studies \cite{Reiss1972, Wang2009, BUICA2008107} showed that multiphoton ionization by linearly polarized light beams is predominant in the higher order interaction regime. These theoretical findings were experimentally confirmed for low \cite{Fox:1971:1416, CERVENAN1974280} and high \cite{PhysRevA.15.1604} order ionization processes. 

A number of Free Electron Laser (FEL) facilities are capable today of producing intense circularly polarized extreme-ultraviolet (XUV) beams \cite{Allaria:2014:041040} which, together with an additional laser, allow ionization of atoms by a circularly polarized XUV + IR light beam. In this process, the XUV pulse pumps the atomic system to several states with well-defined projection of angular momentum, while the IR beam subsequently interacts with an already polarized target. Different total ionization rates are measured and depend on whether the IR beam is co- or counterrotating with the XUV beam. The question of whether the ionization by two co- or counterrotating beams is more efficient depends strongly on the frequency and intensity of the IR beam \cite{Ilchen:2017:013002, Desilva2021}. It was shown that ionization by corotating XUV+IR  beams dominates \cite{Ilchen:2017:013002, Grum:2019:033404} at low IR beam intensities, whereas the result reverses for higher intensities. However, due to the complexity of the process, further investigations need to be carried out to fully understand the process.

In this paper, we will address the two questions mentioned above namely, about the dominance of ionization by linearly or circularly polarized and co- or counterrotating beams, for the case of biharmonic $\omega + 3\omega$ ionization of atoms. Biharmonic beams consist of two co-propagating field components with a fixed phase difference and with frequencies that are integer multiples of the same fundamental frequency $\omega$, i.e.~$m\,\omega + n\,\omega$. In the biharmonic multi-photon ionization of atoms by such beams, the photoelectrons ionized by $n \times (m\,\omega)$- \textrm{or} $m \times (n\,\omega)$-photons are therefore released with the same energy. It has been shown that the interference between the two processes gives rise to properties in photoelectron angular distributions, such as up/down asymmetry \cite{Douguet:2017:105, Gryzlova:2019:063417}, elliptical dichroism \cite{Fifirig_2002, Hofbrucker:2020:3617} or circular dichroism \cite{Volotka_2021}, and it can be even used for the creation and control of electron vortices \cite{Pengel:2017:053003, Kerbstadt:2019:1672583, Armstrong:2019:063416}. In $\omega + \; 2\omega$ ionization of atoms, the photoelectron partial-wave states are orthogonal to each other and, hence, the total ionization rate is independent of interference between the two process \cite{Yian:1992:2353}. However, the interference between the two processes in $\omega + \;2\omega$ ionization of atoms is imprinted in the photoelectron angular distribution~\cite{Prince:2016:176, Giannessi:2018:7774, DiFraia:2019:213904}. In contrast, both photoelectron distributions \textit{as well as} the total rate of $\omega + \;3\omega$ ionization depends on the interference between the one- and three-photon ionization processes, as discussed in the theoretical paper by Chan, Brumer and coworkers~\cite{Brumer:1991:2688}. 

Here we address two main questions in detail. Is it more effective to ionize an atom with a linearly or circularly polarized beams? Are the total electron yields dominant for ionization of atoms by corotating or counterrotating biharmonic fields? We show that the answers to these questions strongly depend on the beam properties and the interference between the one- and three-photon ionization processes. We demonstrate our findings for the ionization of neon $L$ shell electrons. 

This paper is structured as follows. We first introduce our fully relativistic approach, which is based on the third-order perturbation theory and the independent particle approximation in Sec. \ref{Sec.Theory}. Sec.~\ref{Sec.Results} shows our main results, where the total biharmonic $\omega + 3\omega$ ionization rates are compared for various combinations of polarization as an example of ionization of the $2s_{1/2}$ as well as $2p_{3/2}$ electrons of neutral neon atoms. A summary is given in Sec.~\ref{Sec.Summary}. A more detailed description of the first and third-order ionization amplitudes are provided in the Appendix.

\section{Theory}\label{Sec.Theory}

The vector potential of a plane wave photon with frequency $n\omega$, wave vector $\bm{k}$ can be written as
\begin{eqnarray}
    \label{Eq.VectorPotential}
    \bm{A}^{(n\omega)}(\bm{r},t) &=&  \bm{\varepsilon}^{(n\omega)} e^{-i n\omega t + i\bm{k}^{(n\omega)} \cdot \bm{r} }.
\end{eqnarray}
 Here, the polarization is denoted by $\bm{\varepsilon}^{(n\omega)}$ and can be expressed in terms of ellipticity $\gamma^{(n\omega)}$ and basis vectors in helicity representation $\bm{\varepsilon}_{\pm 1}$ as
\small
\begin{equation}
    \label{Eq.PolarizationHelicity}
    \bm{\varepsilon}^{(n\omega)} = \frac{\bm{\varepsilon}_{-1}[1-\gamma^{(n\omega)}]-\bm{\varepsilon}_{+1}[1+\gamma^{(n\omega)}]}{\sqrt{2[1+(\gamma^{(n\omega)})^2]}}.
\end{equation}
\normalsize
\noindent The ellipticity takes values in the range ${|\gamma^{(n\omega)}| \leq 1}$, where ${\gamma^{(n\omega)}=-1}$ refers to left-circularly, ${\gamma^{(n\omega)}=0}$ to linearly and ${\gamma^{(n\omega)}=1}$ to right-circularly polarized light.

Let us moreover consider a biharmonic $\omega + 3\omega$ beam that consists of a fundamental frequency $\omega$ and its third harmonic-order co-propagating along the quantization axis $\bm{\hat{k}} || \bm{\hat{z}}$. The vector potential of such a biharmonic beam can be written as
\begin{eqnarray}
    \label{Eq.BiharmonicField}
    \bm{A}(\bm{r},t) &=& A_0^{(\omega)}\bm{A}^{(\omega)}(\bm{r},t) + e^{i\Phi}A_0^{(3\omega)}\bm{A}^{(3\omega)}(\bm{r},t),
\end{eqnarray}
where $\Phi$ defines a constant phase shift between the two beam components. Furthermore, $A_0^{(n\omega)}$ is the amplitude of the vector potential of each component and is directly proportional to the flux of the component ${F^{(n \omega)} = (A_0^{(n\omega)})^2}$ and its intensity ${I^{(n \omega)} = n \omega F^{(n \omega)}}$.

\begin{figure}
    \centering
    \includegraphics[width = 0.50\textwidth]{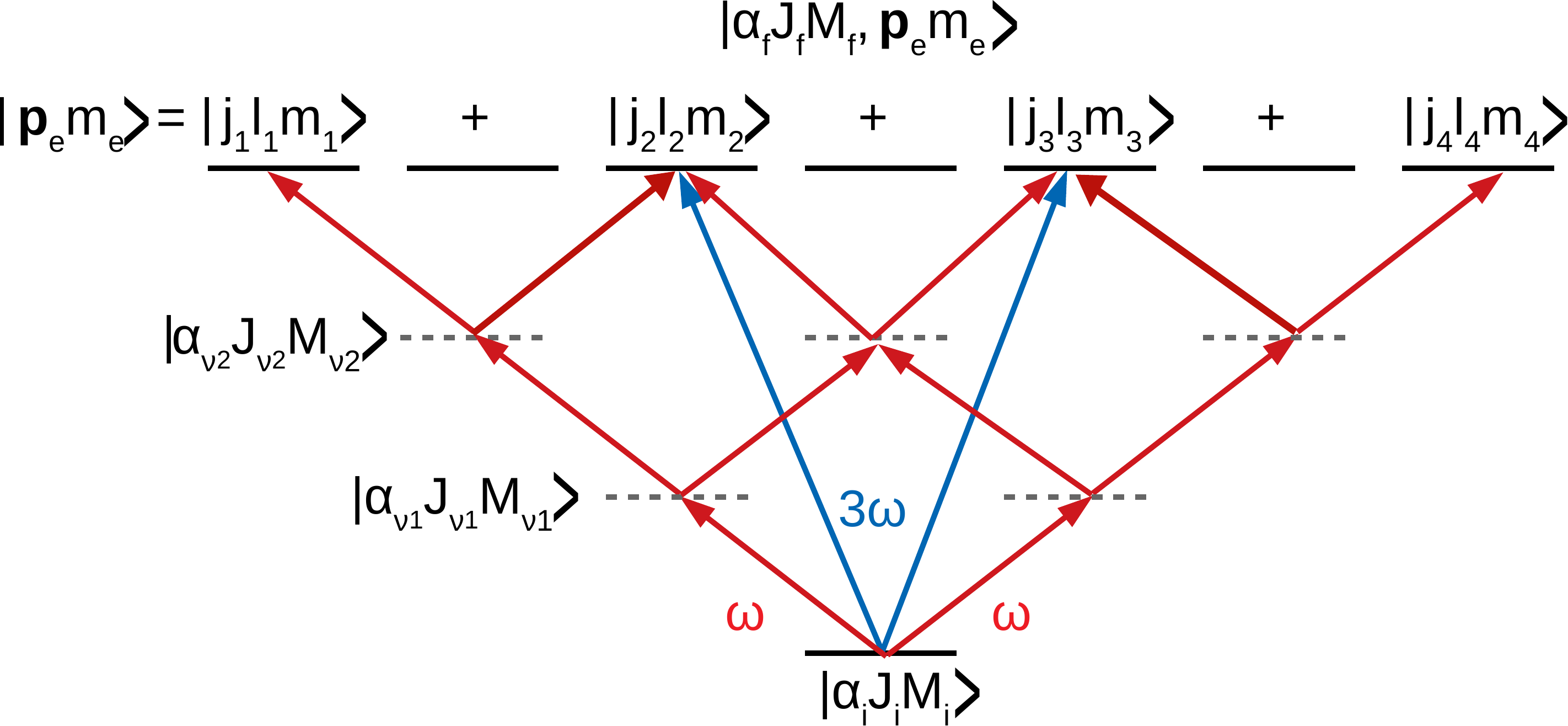}
    \caption{ Schematic representation of the electric dipole ionization pathways in biharmonic $\omega + 3\omega$ ionization of atoms. Both the one- and three-photon ionization always lead to at least one partial wave of the free electron that coincides and that gives rise to the interference effects in the total biharmonic rates.  }
    \label{Fig.Schematic}
\end{figure}
Below, we consider the ionization of a closed-shell neutral atom in an initial many-electron state $\ketm{\alpha_i J_i M_i}$ by a biharmonic $\omega + 3\omega$ field as given in Eq.~(\ref{Eq.BiharmonicField}). The atomic state is characterized by the total angular momentum $J$, its projection $M$ and further quantum numbers $\alpha$ which are necessary to uniquely describe the atomic state. The interaction of the biharmonic field with the atom can lead to a number of different processes. We shall consider the below-threshold ionization ($2\omega < E_b$) of an atom due to the absorption of one photon with energy $3\omega$ or three photons with energy $\omega$, as shown schematically in the Fig.~\ref{Fig.Schematic} for the biharmonic $\omega + 3\omega$ ionization. The final state of the system consists of a singly charged ion, a photoelectron which is denoted by $\ketm{\alpha_f J_f M_f, \bm{p}_e m_e}$, momentum $\bm{p}_e$ and the projection of spin $m_e$. In this work, we will consider infinitely long biharmonic pulses, an assumption that applies well for the biharmonic pulses produced by the current FEL's~\cite{Douguet:2017:105, Gryzlova:2018:013420, DiFraia:2019:213904}. We describe the one- and three-photon ionization processes within the lowest-order perturbation theory, whose transition amplitudes are

\small
\begin{equation}
    \label{Eq.TransitionAmplitudeOPI}
    M_{M_i M_f m_e}^{(3\omega)}= \mem{\alpha_f J_f M_f, \bm{p}_e m_e}{\bm{\alpha \cdot} \bm{A}^{(3\omega)}}{\alpha_i J_i M_i},
\end{equation}
\normalsize

%
%

\small
\begin{eqnarray}
    \label{Eq.TransitionAmplitudeTPI}
M_{M_i M_f m_e}^{(\omega)} &=& \SumInt_{\nu_2} \mem{\alpha_f J_f M_f, \bm{p}_e m_e}{\bm{\alpha \cdot} \bm{A}^{(\omega)}}{\alpha_{\nu_2} J_{\nu_2} M_{\nu_2}}  \nonumber \\  \nonumber
        &\times&  \SumInt_{\nu_1} \mem{\alpha_{\nu_2} J_{\nu_2} M_{\nu_2}}{\bm{\alpha \cdot} \bm{A}^{(\omega)}}{\alpha_{\nu_1} J_{\nu_1} M_{\nu_1}} \\
                &\times&  \frac{\mem{\alpha_{\nu_1} J_{\nu_1} M_{\nu_1}}{\bm{\alpha \cdot} \bm{A}^{(\omega)}}{\alpha_i J_i M_i}}{(E_{i}+2\omega-E_{\nu_2})(E_{i}+\omega-E_{\nu_1})}, 
\end{eqnarray}
\normalsize

\noindent respectively, and where $\bm{\alpha}$ denotes the vector of Dirac matrices. Moreover, we make use of the independent-particle approximation, where the electron wave function is represented by a single active electron, while all other electrons are taken into account by a screening potential in the Hamiltonian of the Dirac equation; see~\cite{Hofbrucker:2016:063412} for a detailed description of the numerical method. However, numerical calculations for a many-electron atomic system can be performed with atomic structure theory codes such as JAC~\cite{JAC}. Due to the interaction of the atom with the electromagnetic field, the active electron of the substate described by the principal $n_a$, relativistic $\kappa_a$ quantum numbers, and the projection of total angular momentum $m_a$ given by $\ketm{a} \equiv \ketm{n_a \kappa_a m_a}$ of the atom is promoted into the continuum, leaving a vacancy in the atomic substate. The relativistic quantum number $\kappa_a$ can be obtained from the total ($j_a$) and orbital ($l_a$) angular momentum quantum numbers by $\kappa = (-1)^{l_a+j_a+1/2} (j_a+1/2)$. In the second quantization, the final many-electron state can be described by a Slater determinant wave function with the use of the electron creation $a^\dagger_{\bm{p}_e m_e}$, annihilation operators $a_{n_a \kappa_a m_a}$ and the Clebsch-Gordan coefficients $\sprm{.., ..}{..}$ as 
\begin{eqnarray}
    \label{Eq.FinalStateManyElectron}
\left\vert{\alpha_f J_f M_f, \bm{p}_e m_e}\right\rangle &=& \sum_{m_a M}\sprm{j_a -m_a, J_i M}{J_f M_f} \\ \nonumber
 &\times& (-1)^{j_a - m_a} a^\dagger_{\bm{p}_e m_e} a_{n_a \kappa_a m_a} \left\vert{\alpha_i J_i M}\right\rangle.
\end{eqnarray}
In the independent particle approximation, the many-electron amplitudes (\ref{Eq.TransitionAmplitudeOPI}) and (\ref{Eq.TransitionAmplitudeTPI}) can be simplified to amplitudes which depend only on one-electron wave functions of the active electron,

\small
\begin{eqnarray}
    \label{Eq.TransitionAmplitudeOPISE}
    M_{M_i M_f m_e}^{(3\omega)} &=& \sum_{m_a} \sprm{j_a -m_a, J_i M_i}{J_f M_f} (-1)^{j_a - m_a} \\\nonumber
            &\times& \mem{\bm{p}_e m_e}{\bm{\alpha \cdot} \bm{A}^{(3\omega)}}{a},
\end{eqnarray}

\noindent and
\begin{eqnarray}
    \label{Eq.TransitionAmplitudeTPISE}
    M_{M_i M_f m_e}^{(\omega)} &=&  \sum_{m_a} \sprm{j_a -m_a, J_i M_i}{J_f M_f} (-1)^{j_a - m_a}  \nonumber\\\nonumber
        &\times& \sum_{n_1n_2} \frac{
        \mem{\bm{p}_e m_e}{\bm{\alpha \cdot} \bm{A}^{(\omega)}}{n_2}
        \mem{n_2}{\bm{\alpha \cdot} \bm{A}^{(\omega)}}{n_1}}
        {(E_{a}+2\omega-E_{n_2})(E_{a}+\omega-E_{n_1})}\\ &\times&\mem{n_1}{\bm{\alpha \cdot} \bm{A}^{(\omega)}}{a}.
\end{eqnarray}
\normalsize
To calculate the third-order transition amplitude, we perform two summations over the complete energy spectra of single-electron intermediate states $\ketm{n_{1,2}}$. For the sake of numerical evaluation, it is convenient to expand the transition amplitudes. To do that, we express the photoelectron wave function into its partial-wave components 

\begin{eqnarray}
    \label{Eq.PartialWaveExpansion}
    \ketm{\bm{p}_e m_e} &= & \frac{1}{\sqrt{E_e |\bm{p}_e|}} \sum_{j m_j} \sum_{l m_l} i^l e^{-i{\delta_{\kappa}}} \sprm{l m_l, 1/2 m_e}{j m_j} \nonumber \\ &\times& \ketm{E_e \kappa m_j} Y^{\ast}_{l m_l}(\theta, \phi),
\end{eqnarray}
where the electron energy is given by $E_e = \sqrt{\bm{p}_e^2+1}$, the phases of the partial waves by $\delta_{\kappa}$ and the emission direction of each partial wave in terms of the polar $\theta$ and azimuthal $\phi$ angles by the spherical harmonics $Y_{l m_l}(\theta, \phi)$. Furthermore, the vector potential ${{\bm{A}}^{(n\omega)}}$ can be decomposed into spherical tensors with electric ($p=1$) and magnetic ($p=0$) components of multipolarity $J$ using
\begin{equation}
    \label{Eq.MultipoleExpansion}
{{\bm{A}}^{(n\omega)}} = 4 \pi \sum_{J M p} i^{J-p} [\bm{\varepsilon}^{(n\omega)} \cdot \bm{Y}^{(p)\ast}_{J M} (\hat{\bm{k}})] {\bm{a}_{J M}^ {(p)}} (\bm{r}).
\end{equation}
The explicit form of the two transition amplitudes in Eqs.~(\ref{Eq.TransitionAmplitudeOPISE}) and (\ref{Eq.TransitionAmplitudeTPISE}) are provided in the Appendix for the first- and third-order transition amplitudes, respectively.

The amplitudes of the one- and three-photon ionization are applied to construct the biharmonic $\omega + 3\omega$ ionization rate
%
\begin{eqnarray}
    \label{Eq.TransitionRateGeneral}
     \nonumber  W(\gamma^{(\omega)},\gamma^{(3\omega)}) &=& \int d \Omega \frac{1}{[J_i]} \sum_{M_i M_f m_e} |K^{(\omega)} M_{M_i M_f m_e}^{(\omega)} \\&+& K^{(3\omega)} e^{i\Phi}M_{M_i M_f m_e}^{(3\omega)}|^2,
\end{eqnarray}
\normalsize
with $[J_i]=(2J_i+1)$ and with the prefactors for one- and two-photon ionization $K^{(3\omega)}=\sqrt{\frac{4 \alpha \pi^2 F^{(3\omega)}}{3\omega}}$ and $K^{(\omega)}=\frac{4\pi^{2} (\alpha F^{(\omega)})^{3/2}}{\omega^{3/2}}$, respectively. These prefactors are derived from the $S$-matrix formalism, see e.g., \cite{Achiezer:1969}. The equation (\ref{Eq.TransitionRateGeneral}) for the ionization rate contains all relativistic effects and all multipoles of the electron-photon interaction. In practice, however, it is generally sufficient to account only for the dominant electric dipole transitions. Therefore, in the calculations of our results, the electric dipole approximation was applied.


\section{Results}\label{Sec.Results} 
\subsection{Energy dependence of the total ionization rate}

\begin{figure}
    \centering
    \includegraphics[width=0.49 \textwidth]{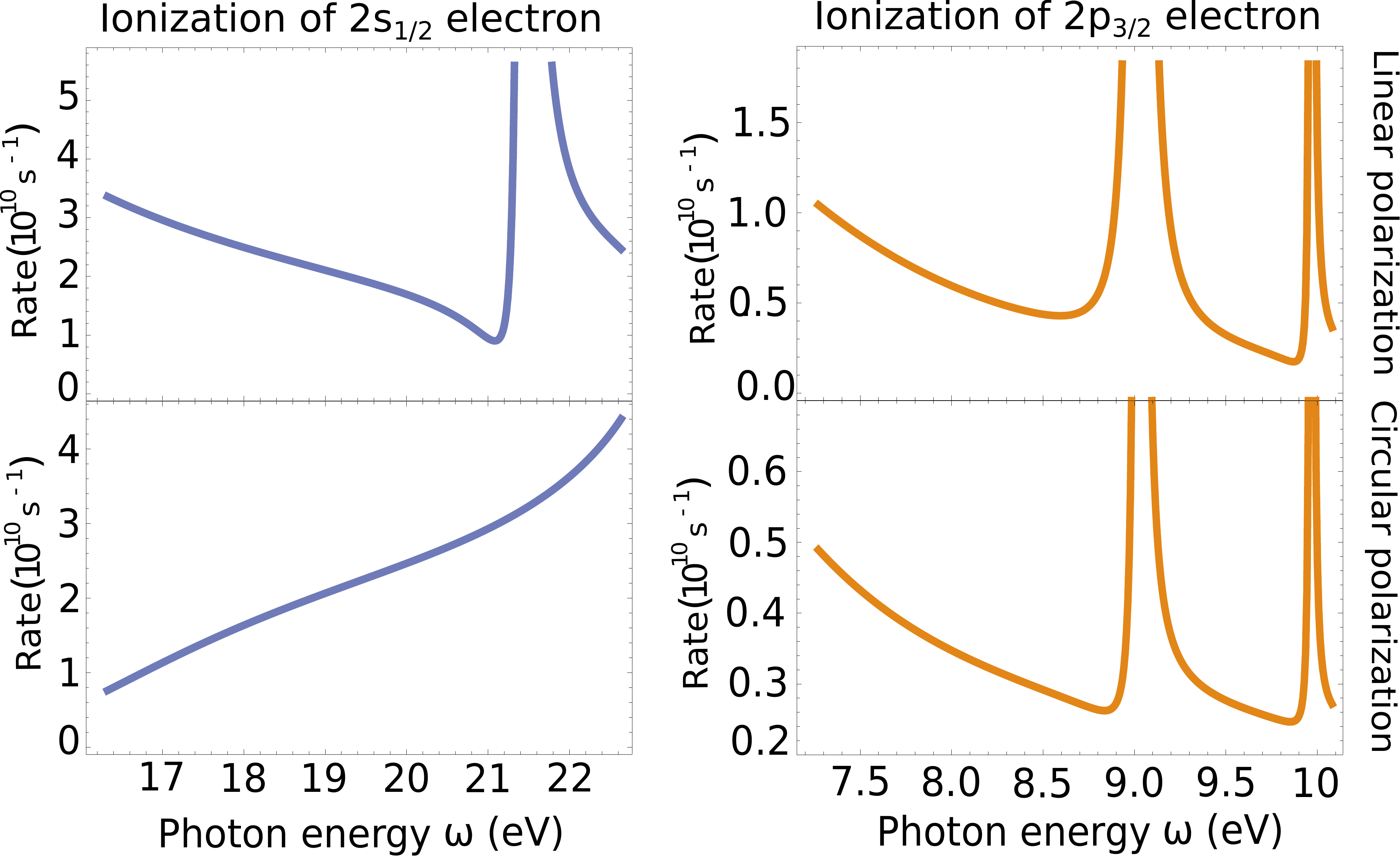}
    \caption{ Total biharmonic $\omega + 3\omega$ ionization rate as a function of incident photon energy. The rates are shown for ionizing the neon $2s_{1/2}$ (left) and $2p_{3/2}$ electrons (right) by linearly (top) and circularly (bottom) polarized biharmonic light are shown. An intensity of the fundamental light beam $ I^{(\omega)} = 1.0 \times 10^{14} \text{W}/\text{cm}^{2}$ was used, while intensity of third harmonic beam was chosen to be $ I^{(3\omega)} = 1.1 \times 10^{11} \text{W}/\text{cm}^{2}$ for ionization of the $2s_{1/2}$ electrons and $ I^{(3\omega)} = 8 \times 10^{11} \text{W}/\text{cm}^{2}$ for $2p_{3/2}$ electrons.  The phase difference between the fundamental and third harmonic light beam was set to zero in the above plots.}
    \label{Fig.Totalrates}
\end{figure}

\begin{figure}
    \centering
    \includegraphics[width=0.46 \textwidth]{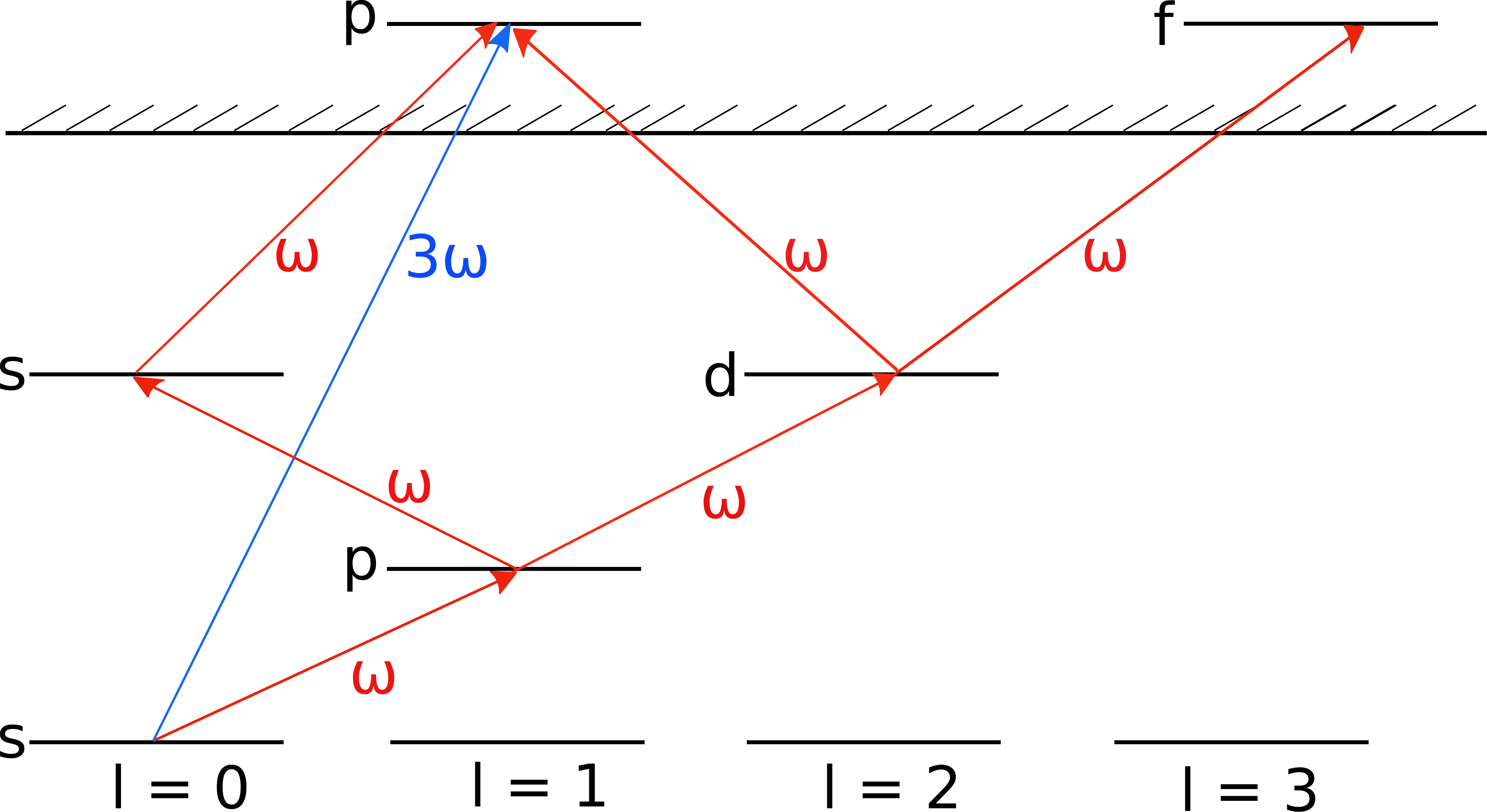}
    \caption{Schematic representation of the electric dipole ionization pathways in biharmonic $\omega +3 \omega$ ionization of atoms for a specific case of s electron. Both the one- and three-photon ionization pathways leads to $s\rightarrow p$ ionization pathway which proceeds through different paths. In this example, the one-photon ionization pathway interferes with two three-photon ionization pathways.}
    \label{Fig.newfigure}
\end{figure}

The total ionization rate of the biharmonic $\omega + 3\omega$ ionization of atoms can be divided into the one- and three-photon ionization rates and their interference. The interference includes all combinations of the ionization pathways of the two processes, the angular dependence of which is determined by the spherical harmonics; see Eq.~(\ref{Eq.PartialWaveExpansion}). Owing to the angular integration in Eq.~(\ref{Eq.TransitionRateGeneral}) and orthogonality of spherical harmonics, only partial waves with the same orbital angular momentum interfere in the total ionization rate. For example, the one-photon ionization of an $s$ electron leads to a $s \rightarrow p$ ionization pathway which comprises both, the $p_{1/2}$ and $p_{3/2}$ partial waves, while three-photon ionization proceeds through the $s \rightarrow p \rightarrow s \rightarrow p$, $s \rightarrow p \rightarrow d \rightarrow p$, $s \rightarrow p \rightarrow d \rightarrow f$ ionization pathways with all the associated fine structure levels, as shown in~Fig.~\ref{Fig.newfigure}. The one-photon ionization pathway ($s \rightarrow p$) in this example interferes only with the first two three-photon ionization pathways ($s \rightarrow p \rightarrow s \rightarrow p$ and $s \rightarrow p \rightarrow d \rightarrow p$), as all of them lead to a final partial wave $p$. This interference in the total biharmonic $\omega + 3\omega$ ionization rate is, however independent of the phases of the photoelectron partial waves $\delta_\kappa$. However, the dependence on the phase difference $\Phi$ between the two beam components remains and is proportional to $\cos \Phi$.

Figure~\ref{Fig.Totalrates} displays the total ionization rate as a function of incident photon energy for biharmonic $\omega + 3\omega$ ionization of neon $2s_{1/2}$ (left) and $2p_{3/2}$ (right) electrons by linearly (top) and co-rotating circularly (bottom) polarized beam. In this figure, the intensity of the the fundamental frequency beam is $I(\omega)=10^{14}$W/cm$^{2}$ and the intensity of the third harmonic beam was chosen to be $I(3\omega)=1.1 \times 10^{11}$W/cm$^{2}$ for ionization of the $2s_{1/2}$ electrons and $I(3\omega)=8\times 10^{11}$W/cm$^{2}$ for ionization of the $2p_{3/2}$ electrons, such that the ionization rates of both processes at threshold energies are comparable. In this work, we use intensity $I(\omega)$ of fundamental and $I(3\omega)$ third-harmonic light field of the order of $10^{14}$W/cm$^{2}$ and $10^{11}$W/cm$^{2}$, respectively, which results in comparable ionization rates for the one- and three-photon ionization processes. Although, it is currently challenging to reach such beam configurations, experiments with similar parameters have been recently successfully carried out~\cite{DiFraia:2019:213904} and advancements in FEL science will likely deliver light beams considered here in near future. The phase difference between the two beam components was set to zero, i.e. $\Phi = 0$, which results in the maximum absolute value for the interference term. The ionization rate for one-photon ionization follows the typical $(3\omega)^{-7/2}$ dependence on the incident photon energy, and hence the dynamical dependence of the biharmonic rate on the incident photon energy pictured in Fig.~\ref{Fig.Totalrates} is dominantly determined by the three-photon ionization process. 

\par In the top left plot, for the biharmonic $\omega + 3\omega$ ionization of the $2s_{1/2}$ electron by linearly polarized light, a local minimum in the rate as well as a resonance behavior appear. This minimum arises from the destructive interference of the one- and three-photon ionization paths and vanishes if the interference is zero, i.e. if the phase difference between the beam components is chosen to be $\Phi=\pi/2$. The resonance in the ionization rate at $\omega=21.5$~eV arises from the $2\omega$ energy matching the energy difference between the ground state of neon and the $1s^2 2s 2p^6 3s$ excited state. No such resonant enhancement of the rate is observed for the biharmonic $\omega + 3\omega$ ionization of the $2s_{1/2}$ electron by circularly polarized beams. This behavior can be readily understood because two circularly polarized photons transfer two units of angular momentum projection to the atom. Since the projection of angular momentum of the excited state is zero, the excitation of the neon in its $^{1}S_{0}$ ground state to the $1s^2 2s 2p^6 3s$ level is forbidden for circularly polarized light. A resonantly increasing ionization rate can be also observed for the biharmonic $\omega + 3\omega$ ionization of the $2p_{3/2}$ electron of neon due to the transition of the ground state neon into the $1s^2 2s^2 2p^5 3p$ (first resonance) and $1s^2 2s^2 2p^5 4p$ (second resonance) excited states. For these transitions, the individual intermediate fine-structure states with different total angular momenta of the excited atom differ by about less than $0.5$~eV in energy, which is not resolved in our calculations. The total rates for biharmonic $\omega + 3\omega$ ionization of the $2p_{1/2}$ electron of neutral neon are similar to those on the right side of Fig.~\ref{Fig.Totalrates}, but shifted by $\approx 0.1$~eV, which arises from the difference in the binding energies of the $2p_{1/2}$ and $2p_{3/2}$ electrons of neon.

\subsection{Circular versus linear polarization in $\omega+3\omega$ ionization}

\begin{figure}[t]
    \centering
    \includegraphics[scale=0.4]{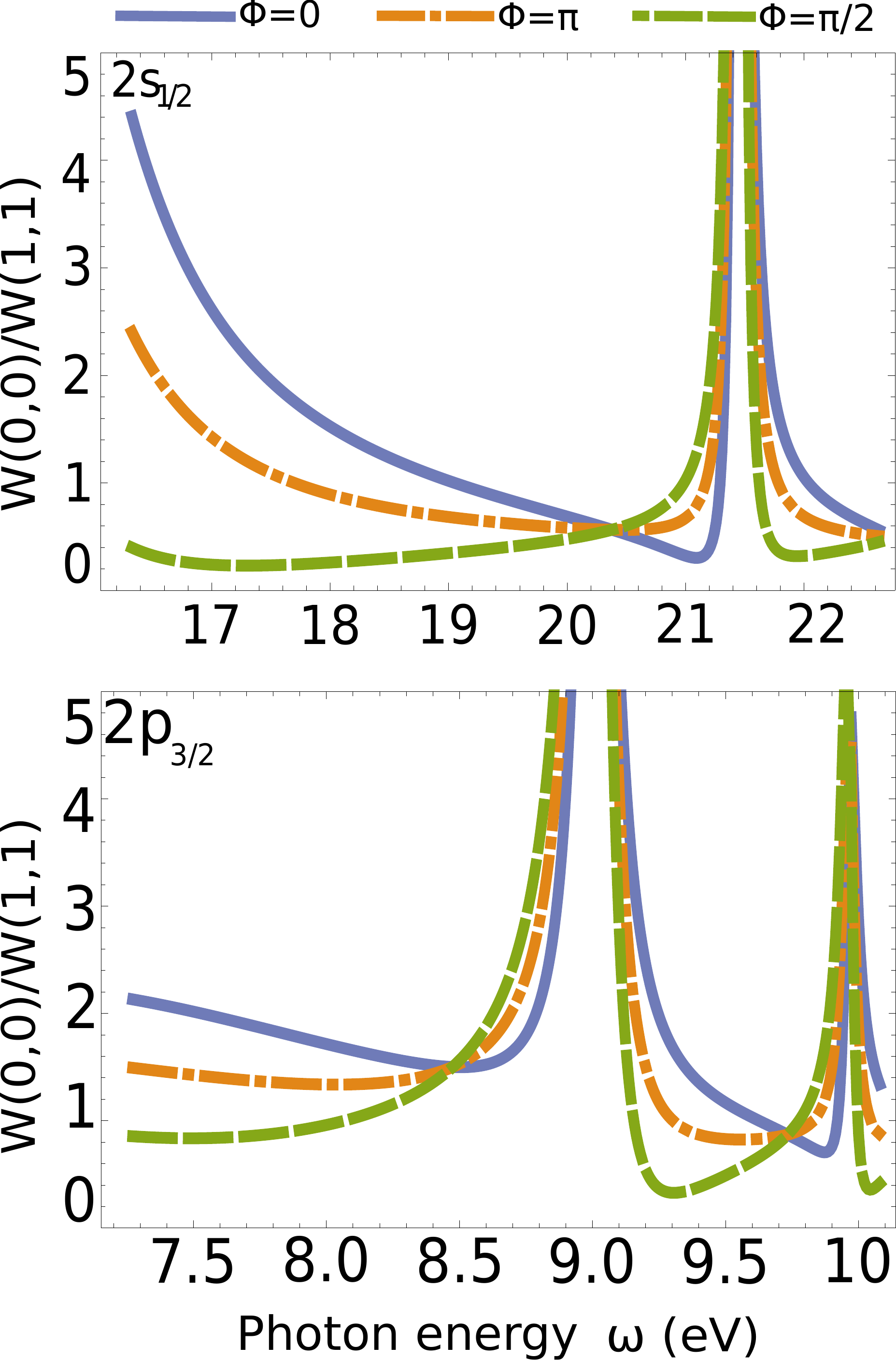}
    \caption{The ratio of the ionization rates for linear polarization \textit{vs.} circular polarization is plotted against the photon energy. The above plot describes the ionization $2s_{1/2}$ (top) electrons and $2p_{3/2}$ (bottom) for three different phase differences between the first and third harmonic light beams. In the above plots, an intensity of fundamental light beam $ I^{(\omega)} = 1.0 \times 10^{14} \text{W}/\text{cm}^{2}$ was applied, while the intensity of third harmonic beam was fixed to be $ I^{(3\omega)} = 1.1 \times 10^{11} \text{W}/\text{cm}^{2}$ for the ionization of the $2s_{1/2}$ electrons and to $ I^{(3\omega)} = 8 \times 10^{11} \text{W}/\text{cm}^{2}$ for $2p_{3/2}$ electrons.  }
    \label{Fig.Ratio_0} 
\end{figure}

In contrast to the total one-photon ionization cross-section, early experiments showed that the multi-photon ionization cross-section depend on the polarization of the ionizing light beam~\cite{Fox:1971:1416, Kogan:1971:1411}. For low-order ionization processes, the cross-section ratio for ionization by circularly and linearly polarized light is well described by the simple estimate $\textrm{max}(\sigma_{\textrm{lin}}/\sigma_{\textrm{circ}}) = N!/(2N-1)!!$ for $N$-photon ionization by Klarsfeld and Maquet \cite{Klarsfeld1972}. This estimate refers to the maximum possible value while the actual ratio might deviate significantly for specific incident photon energies, for example, if the photon energy matches a resonance atomic transition or nonlinear Cooper minimum \cite{Hofbrucker:2019:011401, Hofbrucker:2020:3617}. 

For the biharmonic $\omega + 3\omega$ ionization of atoms, the question of whether ionization by linearly or circularly polarized light dominates the other is more complex than for ionization by monochromatic light, since there are different possible combinations of polarization states that the fundamental and third harmonic can take. In order to facilitate the discussion, we here chose both the fundamental and the third harmonic to have the same polarization, i.e. either they are both linearly or both circularly polarized. The corresponding biharmonic ionization rates are, therefore, represented by $W(0,0)$ and $W(\pm1,\pm1)$, respectively. Furthermore, there is a significant difference between the biharmonic ionization of atoms by linearly and circularly polarized light. The total ionization rate for biharmonic $\omega + 3\omega$ ionization by linearly polarized light comprises the one-, three-photon ionization rates as well as the interference between the corresponding processes. In contrast, the one- and three-photon ionization by circularly polarized light does not lead to any partial wave with the same orbital angular momentum, and hence \textit{do not} interferes in the total ionization rate; see the theory section for more details. 

To evaluate the relative total biharmonic $\omega +3\omega$ ionization rate for ionization by linearly and circularly polarized beams, the ratio $W(0,0)/W(1,1)$ was calculated for three different values of the phase difference between the beam components $\Phi = 0, \pi/2, \pi$ and as a function of incident photon energies. Calculations were performed for the biharmonic $\omega +3\omega$ ionization of the $2s_{1/2}$ and $2p_{3/2}$ electrons of neutral neon and are shown in Fig.~\ref{Fig.Ratio_0}. The results were obtained by using the intensity of the fundamental light beam of $I(\omega) = 1.0 \times 10^{14} \text{W}/\text{cm}^{2}$ and the third harmonic $I(3\omega) = 1.1 \times 10^{11} \text{W}/\text{cm}^{2}$ for the ionization of $2s_{1/2}$ electrons and $I(3\omega) = 8 \times 10^{11} \text{W}/\text{cm}^{2}$ for the ionization of $2p_{3/2}$ electrons. This choice of beam intensities then results in comparable total ionization rates between the one- and three-photon ionization rates for all off-resonant incident beam energies. 

From Fig.~\ref{Fig.Ratio_0}, we can see the behavior of the ratio $W(0,0)/W(1,1)$ as a function of the incident beam energy, which is especially pronounced for the biharmonic ionization of the neon $2s_{1/2}$ electron (upper plot). Let us therefore start with the description of the upper plot. Apart from the enhancement due to the $1s^2 2s 2p^6 3s$ two-photon resonance at around $\omega = 21.5$~eV, the ratio $W(0,0)/W(1,1)$ demonstrates a very strong dependence of the rates on the phase $\Phi$ near the ionization threshold. By choosing the phase $\Phi = 0$ (short-dashed blue), the constructive interference of the one and three-photon ionization is maximized. Therefore, biharmonic ionization by linearly polarized light dominates over ionization by circularly polarized light. For $\Phi = \pi/2$ (long-dashed orange), the interference is zero and $W(0,0)/W(1,1)$ represents purely the ratio of the one- and three-photon ionization rates. For $\Phi=\pi$ (solid red), the interference becomes destructive and hence, the rate corresponding to ionization by linearly polarized light becomes significantly smaller, while rate for ionization by circularly polarized light remains unaffected. 

The sign change of the transition amplitudes near a resonance is often difficult to observe in other processes. However, the sign change of the three-photon ionization amplitudes near to the mentioned two-photon resonance can be read off from the $W(0,0)/W(1,1)$ ratio in the upper plot of Fig.~\ref{Fig.Ratio_0}. For $\Phi = 0$, the interference between the one- and three-photon ionization processes becomes negative, which results in a trough in the $W(0,0)/W(1,1)$ ratio. On the contrary, for $\Phi=\pi$ the resonance enhancement of the ratio appears to be shifted to lower photon energies as a result of the constructive interference, increasing the rates of ionization by linearly polarized light. Similar analysis can be carried out for the ionization of the $p_{3/2}$ electrons, although the described effects become less pronounced. 

\begin{figure}[t]
    \centering
    \includegraphics[width=0.45\textwidth]{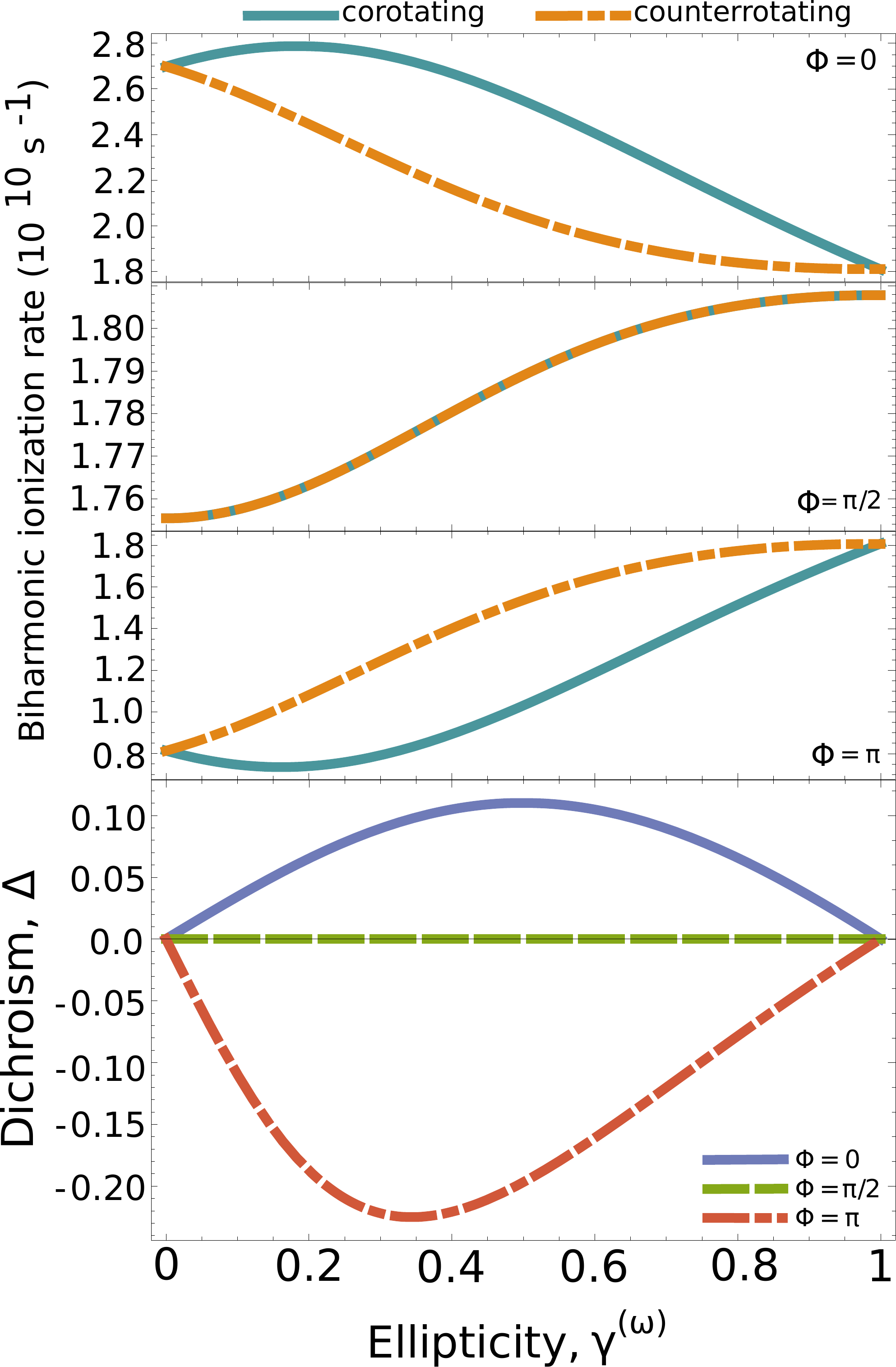}
    \caption{ Ionization of neon $2s_{1/2}$ electrons by co- (solid blue) and counterrotating (dashed-dot orange) biharmonic $\omega + 3 \omega$ beams. The photon energy was chosen near to the threshold at $\omega = 26.7$~eV, the polarization of the third harmonic is chosen to be right-circular $\gamma^{(3\omega)}=1$, while the ellipticity of the fundamental frequency $\gamma^{(\omega)}$ is varied. The intensity of the fundamental beam $I(\omega) = 10^{14}$~W/cm$^2$ and of the third harmonic $I(3\omega)=1.1 \times 10^{11}$~W/cm$^2$ which gives rise to the same one- and three-photon ionization rates. Rates for ionization by biharmonic beams with $\phi = 0, \pi/2, \pi$ phase difference (top three plots, respectively) are shown as functions of the ellipticity of the fundamental beam $\gamma^{(\omega)}$. The dichroism parameter corresponding to these three considerations is shown at the bottom. }
    \label{Fig.Dichroism}
\end{figure}

\subsection{Corotating versus counterrotating beams in $\omega+3\omega$ ionization}

The dominance of ionization by two-color co- or counterrotating beams often refers to circularly polarized light. Since the total ionization rates by circular biharmonic $\omega +3\omega$ beams do not contain any interference terms (as explained before), we will investigate this issue for ionization by right-circularly polarized third harmonic, i.e. $\gamma^{(3\omega)}=1$ and as a function of the ellipticity of the fundamental frequency $\gamma^{(\omega)}$. To enumerate the dominance of co- or counter-rotating beams in biharmonic ionization, we define the dichroism parameter $\Delta(\gamma^{(\omega)})$ as 
\begin{equation}
    \Delta(\gamma^{(\omega)}) = \frac{W(\gamma^{(\omega)},\gamma^{(3\omega)}=1)-W(-\gamma^{(\omega)},\gamma^{(3\omega)}=1)}
                    {W(\gamma^{(\omega)},\gamma^{(3\omega)}=1)+W(-\gamma^{(\omega)},\gamma^{(3\omega)}=1)}.
\end{equation}
In Fig.~\ref{Fig.Dichroism}, we only show the ionization rates for the biharmonic ionization of $2s_{1/2}$ electrons of neon. 

The rates for ionization of neon $2s_{1/2}$ electrons by co- and counterrotatig biharmonic $\omega +3\omega$ beams with different phase difference between the beam components are shown in Fig.~\ref{Fig.Dichroism} as a function of the ellipticity of the fundamental beam together with the corresponding dichroism parameter. This figure shows that the phase difference between the beam components is the key parameter which determines the dominance of one polarization setting over another. For $\Phi=0$ (first plot), the interference between one- and three-photon ionization is constructive, leading to the dominance of the ionization by co-rotating beams. For the phase shift $\Phi=\pi/2$ (second plot), no interference occurs between the two processes, and therefore the total biharmonic rates are independent of the sign of the ellipticity of the fundamental frequency component. For $\Phi=\pi$ (third plot), the interference between the process is destructive which leads to the dominance of the ionization by counterrotating biharmonic beams. The fourth plot shows the dichroism parameters for the three scenarios discussed. The zero dichroism parameter for $\Phi=\pi/2$ reflects the equal rates of ionization by co-rotating and counter-rotating biharmonic beams. Fig.~\ref{Fig.Dichroism} also reveals that the dichroism arising from destructive interference for $\Phi=\pi$ reaches higher absolute values than the dichroism for $\Phi=0$. This can be understood from the comparison of the ionization rates. While the magnitude of the interference is the same for both $\Phi=0$ and $\Phi=\pi$, the latter case results in lower ionization rates which then leads to higher dichroism values. For ionization of $2p_{3/2}$ electrons, the dependence of the dichroism parameter on the ellipticity of the fundamental frequency beam and the phase difference between the beam components $\Phi$ is the same for the ionization of both $2s_{1/2}$ and differs only in magnitude.

\section{Summary}\label{Sec.Summary}

The polarization effects in biharmonic $\omega + 3\omega$ ionization of atoms were studied within the third-order perturbation theory. In particular, the total rates for ionization by linearly and circularly polarized beams were compared, and the dominance of ionization by co- and counterrotating elliptically polarized beams was analyzed. In an example of biharmonic ionization of the neon $L$ shell, we showed that the dominance of a particular beam polarization over another is strongly influenced by interference in the biharmonic ionization process and can be controlled by an appropriate choice of beam parameters. This interference can be controlled most efficiently by varying the phase difference between the biharmonic beam components, which has the strongest effect and has a simple dependence on $\cos(\Phi)$.

\section*{Acknowledgments}
\textbf{Disclaimer:} \textit{The results presented in this paper are based on work performed before February $24^{th}$ 2022}.\\
A.V.V. acknowledges financial support by the Government of the Russian Federation through the ITMO Fellowship and Professorship Program and by the Ministry of Science and Education of the Russian Federation (Project No. 075-15-2021-1349).
\appendix
\section*{Appendix}
\label{Sec.Appendix}
Using the multipole (\ref{Eq.MultipoleExpansion}) as well as the electron partial wave expansion (\ref{Eq.PartialWaveExpansion}) and carrying out the angular integration over the spatial direction $\bm{\hat{r}}$ of the electron wave functions, the transition amplitudes (\ref{Eq.TransitionAmplitudeOPISE}) and (\ref{Eq.TransitionAmplitudeTPISE}) can be written in the following form
\begin{widetext}
\begin{eqnarray}\label{Eq.AmplitudeOPIExpanded}
M_{M_i M_f m_e}^{(3\omega)}&& = 4\pi \sum_{j m_j}\sum_{l m_l} (-i)^l e^{i\delta_{\kappa}} \sprm{l m_l, 1/2m_e}{j m_j} Y_{l m_l}(\bm{\hat{p}}_e) \sum_{J M p} i^{J-p} [\bm{\hat{\varepsilon}}^{(3\omega)} \cdot \bm{Y}_{JM}^{(p)}] \sum_{m_a} \sprm{j m_j, J M}{j_a m_a}  [j]^{-1/2} \nonumber \\
            && \times (-1)^{j_a - m_a} \sprm{j_a -m_a, J_i M_i}{J_f M_f} \redmem{j}{T_{J}}{j_a} U_{\kappa}(pJ)
\end{eqnarray}
and
\begin{eqnarray}
\label{FinalTransitionAmplitude}\label{Eq.AmplitudeTPIExpanded}
M_{M_i M_f m_e}^{(\omega)} &=& \nonumber 
	16 \pi^2 \sum_{jm_j} \sum_{lm_l}(-i)^l e^{i\delta_{\kappa}} \sprm{l m_l,1/2  m_e}{j m_j}Y_{l m_l}(\bm{\hat{p}}_e)
	\sum_{J_1 M_1 p_1 } \sum_{J_2 M_2 p_2} \sum_{J_3 M_3 p_3}
	i^{J_1-p_1+J_2-p_2+J_3-p_3} 
	[\bm{\hat{\varepsilon}}^{(\omega)} \cdot \bm{Y}_{J_1 M_1}^{(p_1)}] \\ \nonumber &\times& 
	[\bm{\hat{\varepsilon}}^{(\omega)} \cdot \bm{Y}_{J_2 M_2 }^{(p_2)}]
	[\bm{\hat{\varepsilon}}^{(\omega)} \cdot \bm{Y}_{J_3 M_3 }^{(p_3)}] 
	\sum_{j_{n_1} l_{n_1} m_{n_1}}\sum_{j_{n_2} l_{n_2} m_{n_2}}
	[j_{n_1},j_{n_2}, j]^{-1/2}\sprm{j m_j,J_3 M_3}{j_{n_2} m_{n_2}}\\ \nonumber &\times& \sprm{j_{n_2} m_{n_2},J_2 M_2}{j_{n_1} m_{n_1}}\sum_{m_a} \sprm{j_{n_1} m_{n_1}, J_1 M_1}{j_a m_a} \sprm{j_a -m_a, J_i M_i}{J_f M_f} 
	\\ 
	&\times& (-1)^{j_a - m_a}  \redmem{j}{T_{J_3}}{j_{n_2}}\redmem{j_{n_2}}{T_{J_2}}{j_{n_1}} \redmem{j_{n_1}}{T_{J_1}}{j_a}U^{(\kappa_{n_2},\kappa_{n_1})}_{\kappa}(p_1 J_1, p_2 J_2, p_3 J_3),
\end{eqnarray}
in terms of radial transition amplitudes for one-photon
\begin{equation}
    \label{Eq.TransitionAmplitudeRadialOPI}
    U_{\kappa}(p J)= R_{\kappa \kappa_a}(p J)
\end{equation}
and three-photon ionization
\begin{equation}
    \label{Eq.TransitionAmplitudeRadialTPI}
    U^{(\kappa_{n_2},\kappa_{n_1})}_{\kappa}(p_1 J_1, p_2 J_2, p_3 J_3)=\SumInt_{n_1} \SumInt_{n_2} \frac
        {R_{\kappa \kappa_{n_2}}(p_3J_3)R_{\kappa_{n_2} \kappa_{n_1}}(p_2 J_2)R_{\kappa_{n_1} \kappa_a}(p_1 J_1)}
        {(\epsilon_{n_a \kappa_a}+2\omega-\epsilon_{n_{n_2} \kappa_{n_2}})(\epsilon_{n_a \kappa_a}+\omega-\epsilon_{n_{n_1} \kappa_{n_1}})}.
\end{equation}

\end{widetext}

The transition amplitudes $U_{\kappa}(p J)$ and $U^{(\kappa_{n_2}, \kappa_{n_1})}_{\kappa}(p_1 J_1, p_2 J_2, p_3 J_3)$ of course depend on the principal quantum numbers of each involved electronic state, however, this dependence was left out from the notation for practical purposes. The angular integration of the space coordinate is given by

\small
\begin{equation}
\redmem{j_f}{T_{J}}{j_i}=(-1)^{j_i+j_f-J+1}[j_i]^{1/2} \sprm{j_i 1/2, J 0}{j_f 1/2} \Pi_{l_i, l, J},
\end{equation}
\normalsize
where $\Pi_{l_i, l, J} = 1$ if $l_i+l+J$ is even and $\Pi_{l_i, l, J} = 0$ otherwise.
\noindent In the transverse (velocity) gauge, the radial integrals are explicitly given for the magnetic  ($p=0$, or $pJ = M J$) transitions
\begin{eqnarray}
 R_{\kappa_f \kappa_i}(M J) &=& i\sqrt{\frac{[J](J+1)}{4J\pi}} \int _ { 0 } ^ { \infty } d r \frac { \kappa _ { i } + \kappa _ { f } } { J + 1 } j _ { J } ( k r ) \nonumber \\ 
 &\times& \big[ P _ { i } ( r ) Q _ { f } ( r ) + Q _ { i } ( r ) P _ { f } ( r ) \big], 
\end{eqnarray}
where $j_J(x)$ are the spherical Bessel functions, the radial wave functions $P(r)$ and $Q(r)$ are the large and small components of the radial Dirac wave functions for the orbital with principal and Dirac quantum numbers $n_i$ and $\kappa_i$, respectively. These components are obtained from single-electron Dirac equation, with a screening potential in the Hamiltonian, which partially accounts for the interelectronic interaction. We compared a number of different potential models. The core-Hartree potential, which reproduces the binding energies in good agreement with the experimental values, was used to produce the results presented in this work. For the electric transitions ($p=1$, or $pJ = E J$)
\begin{eqnarray}
 R_{\kappa_f \kappa_i}(E J) &=& i\sqrt{\frac{[J](J+1)}{4J\pi}}
 \int _ { 0 } ^ { \infty } d r \Bigg\{ - \frac { \kappa _ { i } - \kappa _ { f } } { J + 1 } 
 \big[ j _ { J } ^ { \prime } ( k r )  \nonumber\\
 &+&  \frac { j _ { J } ( k r ) } { k r } \big] \big[  P _ { i } ( r ) Q _ { f } ( r ) + Q _ { i } ( r ) P _ { f } ( r ) \big]  \\\nonumber  &+& { J \frac { j _ { J } ( k r ) } { k r } \big[ P _ { i } ( r ) Q _ { f } ( r ) - Q _ { i } ( r ) P _ { f } ( r ) \big] \Bigg\} } .
\end{eqnarray}
In the length gauge, this integral is given by
\begin{eqnarray}
  R_{\kappa_f \kappa_i}(E J) &=&  i\sqrt{\frac{[J](J+1)}{4J\pi}}
  \int _ { 0 } ^ {\infty} dr  ~ j _ { J } ( k r ) 
  \big[ P _ { i } ( r ) P _ { f } ( r )  \nonumber\\  \nonumber
  &+& Q _ { i } ( r ) Q _ { f } ( r ) \big] + j _ { J + 1 } ( k r ) \Big\{ \frac { \kappa _ { i } - \kappa _ { f } } { J + 1 }  \\ \nonumber
  &\times&  \big[ P _ { i } ( r ) Q _ { f } ( r ) Q _ { i } ( r ) P _ { f } ( r ) \big] + \big[ P _ { i } ( r ) Q _ { f } ( r ) \\ 
  &-& Q _ { i } ( r ) P _ { f } ( r ) \big] \Big\}.
\end{eqnarray}
The presented results were calculated in the velocity gauge, however, the calculations were performed in both velocity and length gauges to check the consistency and accuracy of our calculations.

%


\bibliography{Bibliography}
%
\end{document}